\documentclass[useAMs,a4paper]{mn2e}
\usepackage{savesym}
\usepackage{graphicx}
\expandafter\let\csname equation*\endcsname\relax
  \expandafter\let\csname endequation*\endcsname\relax 
\usepackage{subfig}
\usepackage{amsmath}
\usepackage{amssymb}
\usepackage{verbatim}
\usepackage[yyyymmdd,hhmmss]{datetime}
\usepackage{array}
\usepackage{times}
\usepackage[total={17.8cm,24.0cm},centering]{geometry} 
\usepackage{color}

\newcommand{\be}{\begin{equation}}
\newcommand{\beq}{\begin{equation}}
\newcommand{\ee}{\end{equation}}
\newcommand{\eeq}{\end{equation}}
\newcommand{\eea}{\end{eqnarray}}
\newcommand{\bea}{\begin{eqnarray}}

\newcommand{\dd}{\partial}

\newcommand\W {{W^r_{\ \phi}}}

\title[Evolution of Kerr Discs and TDEs ]{The evolution of Kerr discs and late-time tidal disruption event light curves}
\author [Steven A. Balbus, Andrew Mummery]{Steven A. Balbus\thanks{E-mail:
steven.balbus@physics.ox.ac.uk}, Andrew Mummery
\\
Oxford Astrophysics, Denys Wilkinson Building, Keble Road, Oxford, OX1 3RH, United Kingdom}
\begin{document}

\date{}

\pagerange{\pageref{firstpage}--\pageref{lastpage}} \pubyear{2018}

\maketitle

\label{firstpage}

\begin{abstract} 
An encounter between a passing star and a massive black hole at the centre of a galaxy, a so-called tidal disruption event or TDE,  may leave a debris disc that subsequently accretes onto the hole.  We solve for the time evolution of such a TDE disc, making use of an evolutionary equation valid for both the Newtonian and Kerr regimes.    The late time luminosity emergent from such a disc is of interest as a model diagnostic, as it tends to follow a power law decline.  The original simple ballistic fallback model, with equal mass in equal energy intervals, produces a $-5/3$ power law, while standard viscous disc descriptions yield a somewhat more shallow decline, with an index closer to $-1.2$.   Of four recent, well-observed tidal disruption event candidates however, all had fall-off power law indices smaller than 1 in magnitude.    In this work, we revisit the problem of thin disc evolution, solving this reduced problem in full general relativity. 
Our solutions produce power law indices that are in much better accord with observations.   The late time observational data from many TDEs are generally supportive, not only of disc accretion models, but of finite stress persisting down to the innermost stable circular orbit.  
\end{abstract}

\noindent
Complied at \today\ \currenttime\ .

\section{Introduction}

The evolution of thin accretion discs is generally studied via techniques first developed by Lynden-Bell \& Pringle (1974, hereafter LBP; Pringle 1981 for a review).   They showed that an evolving thin Keplerian disc, subject to a viscous torque, obeyed a simple diffusion-like equation.   
The classical (Newtonian) disc evolutionary equation has found practical use in many astrophysical systems, including dwarf nova eruptions (Frank et al. 2002), protoplanetary discs (Fromang, Balbus, \& Terquem 2002) and, the topic of this paper, tidal disruption events (hereafter TDEs) (Cannizzo, Lee, \& Goodman 1990).   While {\em equilibrium} relativistic thin disc theory was developed nearly half a century ago, the ability to study relativistic disc evolution by the methods similar to those introduced by LBP is only now being developed.    (Riffert [2000]  also examined the spreading of viscous discs in the relativistic regime, but for reasons that we will discuss in \S2.4.2 below, out treatment differs somewhat from this.)   As an alternative to the much more expensive direct numerical simulation, the one-dimensional relativistic equation promises to be a practical theoretical tool, occupying the middle ground between computational rigour and phenomenological modelling.  

In one of the appendicies of a paper studying magnetic stresses in accretion discs, Eardley \& Lightman (1975) first presented the form of this evolutionary appropriate for Kerr spacetime\footnote{We are grateful to P.\ Ivanov for drawing this paper to our attention.}.  Following the Novikov \& Thorne (1974) and Page \& Thorne (1974) equilibrium models, the viscosity (in reality turbulent transport) is represented by an anomalous stress tensor.    First invoked by Shakura \& Sunyaev (1973), the anomalous stress both enhances the transport of angular momentum and drives significant dissipative heating.   Emerging in the form of radiative losses, the latter energy source is the mechanism by which X-ray emitting accretion discs are ultimately observed. 

Very little was done by way of investigating explicit solutions of the Eardley-Lightman equation until very recently, when Balbus (2017), unaware of the earlier derivation, rediscovered the Kerr evolutionary equation (in slightly different coordinates), and presented formal WKB modal solutions for both finite and vanishing stress inner disc bondary conditions.  
In the current work,  we examine solutions of the Kerr disc equation more generally, with a focus on the bolometric light curves, and compare their late time evolution with TDEs (Rees 1988).   Our goal is to understand some pertinent but puzzling features suggested by the observational data.    A recent compilation by Auchettl, Guillochon, \& Ramirez-Ruiz (2017; hereafter AGR) is revealing.   At late times, TDEs are expected to display light curves $L(t)$ that vary as a power law $n$ in time $t$, $L\sim t^n$.  The original Rees (1988) ``fallback'' model, which assumed equal mass in equal energy intervals (and ballistic dynamics), leads to $n=-5/3$.    Disc accretion models (Cannizzo et al.\ 1990) extend the duration of the emission somewhat, with a typical index of $n\simeq-1.2$.  However, the late time AGR power law indices of confirmed X-ray TDEs clustered around $n\simeq -0.75$, a much more shallow fall-off.      There is no widely accepted explanation of how such an X-ray index might arise.   We put forth here a simple and rather surprising solution to this puzzle: something very close to this value is expected from a population of time dependent accreting Keplerian discs.    The difference between our findings and those of Cannizzo et al. (1990) arises from our technique of smoothly joining an outer Keplerian disc solution to an inner relativistic disc, which then terminates at an innermost stable circular orbit (ISCO),  {\em which may have finite stress.}   The presence of this ISCO boundary zone is itself hardly involved with the {\em direct} production of the light curve---it is the external Keplerian zone that is responsible for the bulk of the observed luminosity.   Nevertheless, the ISCO boundary condition is important because it results in a slightly different admixture of external Keplerian modes (exponentially declining Laplace transforms) compared with the pure Newtonian disc.   The altered admixture in turn alters the late time dependence of the disc's thermal emission, ultimately leading to a less steep decline, a sort of feedback effect.   Encouragingly, this also is what observations seem to show.      

Note that the solutions we describe here in principle allow for a measurement of the stress present at the ISCO: the outer Keplerian combination of modes, and thus the late time luminosity, depend rather sensitively upon whether the stress vanishes at the ISCO radius or not.   We find that the imposition of a vanishing stress tensor leads to a much steeper fall-off in $L(t)$, one that replicates the late time behaviour seen in the Cannizzo et al. (1990) calculations.   This, however, accords less well with observations.   A nonvanishing stress condition at the ISCO, by contrast, tips the outer modal balance and leads to a more shallow fall-off for $L(t)$, in better accord with the data.   This interesting point, and how it relates to the stability arguments in Balbus (2017) (which advanced a vanishing stress condition!), is discussed more fully in \S 3.    The question of finite versus vanishing stress at the ISCO still remains a point of contention within the disc community.

The plan of this paper is as follows.    In \S 2, we lay out the fundamental solution to the Kerr evolutionary equation.   We first present the results of direct numerical integration, both for the case of finite stress at the ISCO as well as for vanishing stress.  The late time luminosity behaviour is found to be well fit by a declining power law in both cases, but with a much steeper fall-off for the vanishing ISCO stress.   The finite stress case appears to be compatible with observed results of confirmed TDEs.  The numerical integration is followed by an analysis of Laplacian-Bessel normal modes of the disc system.   With the appropriate ISCO boundary conditions, it is possible to understand why the two different late time behaviours arise.   Finally in \S 3 we discuss the observational implications of our findings, note the limitations of our simple model, and suggest further developments motivated by the present encouraging study.   The Appendices contain technical mathematical details pertinent both to the numerical and analytic discussions.     
  
We observe the same notational conventions of Balbus (2017).  {\em The speed of light is set to unity throughout.}   Greek indices $\alpha, \beta, \gamma...$
generally denote spacetime coordinates.   The exception is $\phi$, which is reserved exclusively for the azimuthal angular coordinate.  The time coordinate is labelled $0$.  The metric in local inertial coordinates is
$
g_{\alpha\beta} \rightarrow \eta_{\alpha\beta} = {\rm diag\ }(-1,1,1,1).
$
Other notation is standard: $G$ is the gravitational constant, $M$ the central black hole mass, $J$ the central black hole angular momentum, $a = J/M$ the black hole spin parameter, and $r_g=GM$ the gravitational radius.

\section{Analysis}

\subsection{Governing equation}
Our coordinates are cylindrical Boyer-Lindquist for a Kerr disc: $r$ (radial), $\phi$ (azimuthal), and $z$ (vertical).   We seek the evolution of the azimuthally-averaged, height-integrated disc surface density $\Sigma (r, t)$.   The contravariant four velocity of the disc fluid is $U^\mu$; the covariant counterpart is $U_\mu$.  The specific angular momentum corresponds to $U_\phi$, a covariant quantity.     There is an anomalous stress tensor present, $\W$, due to low level disk turbulence, which is a measure of the correlation between the fluctuations in $U^r$ and $U_\phi$ (Eardley \& Lightman 1975, Balbus 2017).  This is, as the notation suggests, a mixed tensor.   
As noted earlier in the Introduction, $\W$ serves both to transport angular momentum as well as to extract the free-energy of the disc shear, which is then thermalised and radiated from the disc surface, both assumed to be local processes.    

We shall work with the quantity
\beq\label{yy}
Y\equiv \sqrt{g}\Sigma \W,
\eeq
where $g>0$ is the absolute value of the determinant of the (midplane) Kerr metric tensor $g_{\mu\nu}$.
The governing equation for the evolution of the disc is then given by (Balbus 2017):
\beq\label{fund}
{\dd Y\over \dd t} =  {\W\over U^0}{\dd\ \over \dd r}{1\over U'_\phi} \left[   {\dd Y \over \dd r}- U_\phi U^\phi (\ln\Omega)' Y \right],
\eeq
where the primed notation $'$ denotes an ordinary derivative with respect to $r$.    A direct rendering, or ``relativistic upgrade,'' from the original Newtonian equation (Pringle 1981; Balbus \& Papaloizou 1999), would end with the first term on the right.   The second term is a further relativistic correction stemming from the photon angular momentum loss.    Note that if $\W$ has a functional dependence upon
$\Sigma$, the stress would be implicitly time-dependent.  In that case, equation (\ref{fund}) should be modified to: 
\beq\label{fund2}
{\dd (Y/\W)\over \dd t} =  {1\over U^0}{\dd\ \over \dd r}{1\over U'_\phi} \left[   {\dd Y \over \dd r}- U_\phi U^\phi (\ln\Omega)' Y \right].
\eeq
The metric tensor has disappeared from the evolutionary equation in any explicit form, entering implicitly from the definition of $Y$.  

Before proceeding to solutions of equation (\ref{fund}), let us recall its limitations.  We are of course using thin disc theory, which ignores terms of quadratic
or higher order in the ratio of the disk scale height to radius, $H/r$.   This is equivalent to ignoring pressure terms relative to those involving rotational energy.   Abramowicz et al. (1988) investigated the effects of such terms on the accretion of ``slim discs'' (which assumes small but finite $H/r$), noting their importance near the ISCO.   Here it should be noted that the validity of equation (\ref{fund}) does not depend upon the neglect of pressure  relative to rotation; it depends only on the notion that the concept of some sort of suitably averaged $\W$ turbulent tensor makes sense.  In this respect, it is no more restictive than any reduced theory of disc turbulence, which in essence includes all theoretical modelling of such flow.    Where thin disc theory may be inaccurate is in its handling of the energetics, for which it assumes efficient local radiation.   This is indeed likely to break down near the ISCO.  As we shall see, however, the net luminosity emerging from our solutions appears to be dominated by the inner, but still robustly Keplerian, regions of the disc.   It is therefore useful to develop and understand the predictions of thin disc theory, while granting its omissions and shortcomings, which invite further consideration.

\subsection {Compact formulation and a simple analogue model}

In this paper, we will work entirely with equation (\ref{fund}).   Following Balbus (2017), we define $Q$ by
\beq\label{dQ}
{dQ\over dr} = - U_\phi U^\phi (\ln\Omega)'.
\eeq
Equation (\ref{58Q}) in Appendix A2 shows that for Kerr geometry, we have the useful identity
\beq
e^{-Q} = U^0.
\eeq
Then, with 
\beq
\zeta =Y/U^0,
\eeq
equation (\ref{fund}) becomes 
\beq\label{27q}
{\dd\zeta\over \dd t} =  { \W\over (U^0)^2}{\dd\ \over \dd r}{U^0\over U'_\phi} \left[   {\dd\zeta \over \dd r}\right].
\eeq
While the full dynamical equation (\ref{fund}) may be solved directly by numerical methods (we do so in \S\ref{23} below), it is helpful to have a simplified ``toy model'' that retains essential features of the full problem in a mathematically accessible form that may addressed analytically.    The function $U^0$ reduces to unity in the Newtonian limit, and over the entire domain of interest in Kerr geometry, it is smooth, non-vanishing, and bounded.   The qualitative content of equation (\ref{27q}) may thus be retained by ignoring these functions, in effect setting them equal to unity.   By contrast, $U'_\phi$ vanishes at the ISCO, introducing an apparent singularity into the equation, and must handled with more care.    We shall restrict ourselves to the relatively simple case of Schwarzschild geometry for analytic treatment, as this illustrates many of the salient features of the full Kerr problem in a more tractable setting.    Our model equation thus takes the form:
\beq\label{toy}
{\dd y \over \dd t} =  { \W}{\dd\ \over \dd r}\left[ {1\over U'_\phi}   {\dd y \over \dd r}\right].
\eeq
We use the lowercase $y$ to distinguish our model from the true surface density variable $Y$.

\subsection {Numerical formulation and solution}\label{23}
We begin with a summary of the exact numerical solution of equation (\ref{27q}).   The interested reader will find technical details discussed in Appendix A2.   Some care is needed in handling the numerically singular behaviour near the ISCO, where $U'_\phi$ vanishes.  

The mathematical problem to be solved is the evolution of a very compact Gaussian ring, in effect the Green's function solution.   The ring is initially located at $r_0=15.75\, r_g$, a fiducial tidal radius taken from Rees (1988), $\sim 2.6$ times the ISCO radius of a Schwarzschild hole (see below).   At this location, relativity is by no means negligible.
While the short term evolution depends upon the initial radius chosen, the long term evolution of the extended disc does not.

\begin{figure}
	\centering
           \includegraphics[width=7.5cm,clip=true, trim=0cm 0cm 0cm 0cm]{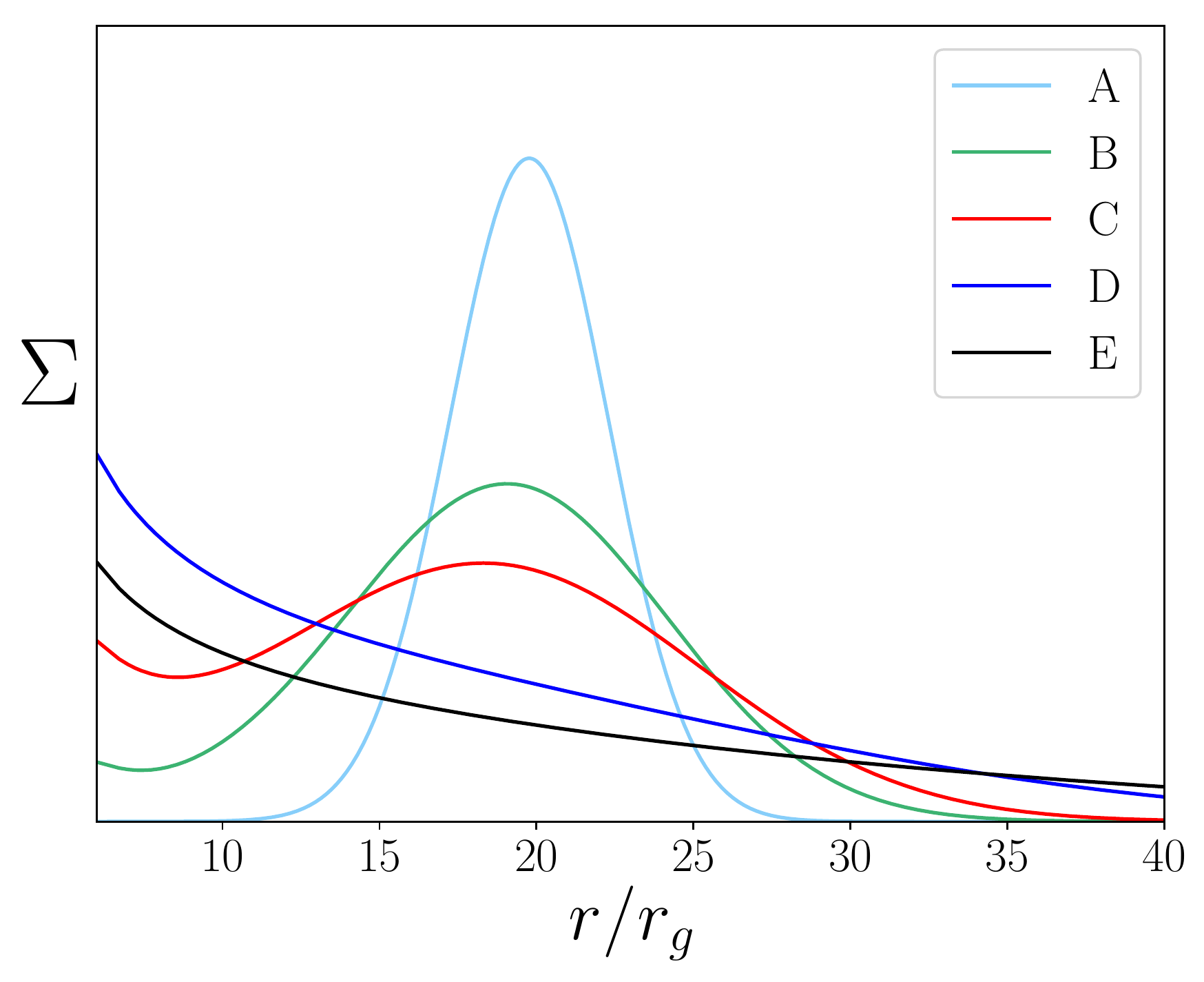}
	\caption{Evolution of the surface density in the Schwarszchild metric. The lines A to E are at progressively later times, showing the initial inwards drift of debris before the ISCO surface density decreases with time. The dimensionless time values (eq. \ref{tau}) for the different stages of evolution are: $\tau = 0.06({\rm A}), 0.24({\rm B}), 0.43({\rm C}), 1.84({\rm D}), 6.06({\rm E})$}
\label{surfacedensity}
\end{figure}

\subsubsection {ISCO boundary conditions}
A representative case is shown in figure (\ref{surfacedensity}).  At early times, the disc spreads both radially inward and outward.  As the inner edge reaches the ISCO, a boundary condition must be specified, the precise nature of which  
depends upon the behaviour of the turbulent stress at the ISCO.   If vanishing stress is imposed, the surface density and thus the local emissivity must also vanish.  
For finite ISCO stress, we show in Appendix A2 that the proper boundary condition is that the radial gradient of $\zeta$ must vanish, as opposed to $\zeta$ itself.   Numerically,  the value of $\zeta$ at the innermost (ISCO) grid point is set equal to its value at the adjacent external grid point. 

\subsubsection {Kerr disc, Finite ISCO stress, $W^r_\phi \propto r^{-1/2}$}
Using equation (\ref{flux}) from Appendix A2, we have evaluated the luminosity profiles $L(t)$ for discs evolving in the Kerr geometry.   for a variety of different black hole spins $a$.    As noted, each disc had the same mass and was initially laid down at $r = 15.75r_g$,  the tidal radius for a solar mass star orbiting a $10^6 M_\circ$ black hole (Rees 1988).  $W^r_\phi \propto r^{-1/2}$ corresponds to a constant viscosity model,
chosen merely as benchmark.   Once the stress at the ISCO is specified, our results are not very sensitive to the precise parameterisation of $\W$.   But the late time behaviour does depend sensitively on whether this stress is finite or vanishing.  

\begin{figure}
\includegraphics[width=.5\textwidth]{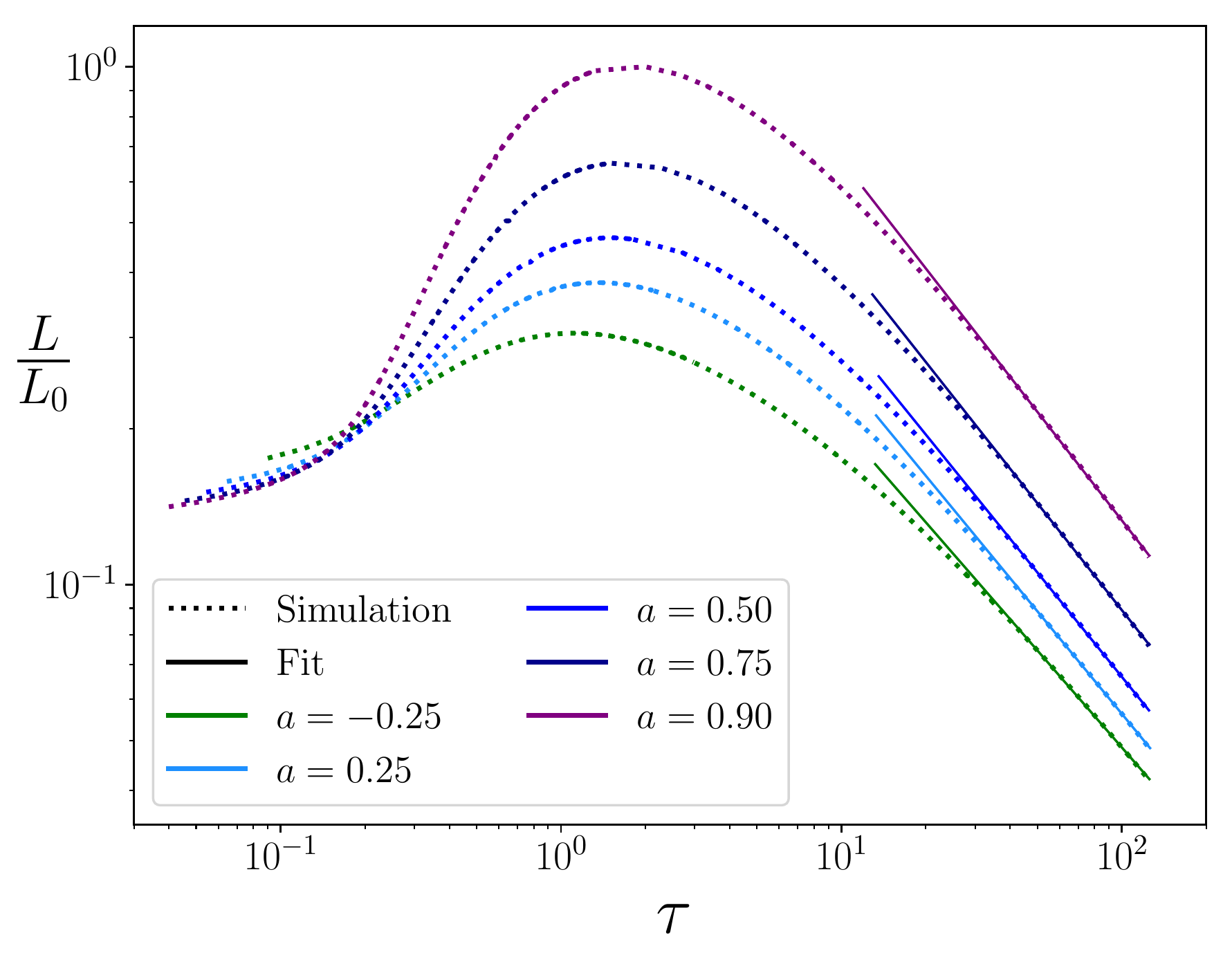}
\caption{Luminosity profiles for a selection of different black hole angular momenta.  $L_0$ is the $a/r_g=0.9$ peak luminosity. The dimensionless time $\tau$ as defined in eq.\  (\ref{tau}) is plotted on the x-axis.  Each profile is well fit by a (straight-line) power law at late times.  See Table 1 for power law values.}
\label{intensityprofiles}
\end{figure}

Even within the restrictions of this simple model, observational TDE emission profile features emerge:  a rapid increase in intensity followed by a gradual monotonic power law decrease in intensity at late times.  The luminosity $L(t) \sim t^n$, with $n$ typically between $-0.6$ and $-0.7$.    Typical fits are shown by the solid lines in Figure (\ref{intensityprofiles}). The best fit values of the decay index, $n$, are shown in Table (\ref{table1}) for several black hole angular momenta.   These solutions are in better accord with late time observations of confirmed TDEs than are classical Newtonian discs with vanishing inner stress.     In particular, the finite stress power law index is always less steep than $t^{-1}$, whereas the Newtonian discs are generally steeper than $t^{-1}$ (Cannizzo et al. 1990).   This is a potentially important result which is developed below.   

\begin{table}
\centering
 \begin{tabular}{||c | c||} 
 \hline
$a/r_g~$ & ~$n~$  \\ [0.5ex] 
 \hline\hline
 $-0.5$ & -0.6 \\ 
 \hline
 $-0.25$&  -0.62\\
 \hline
 $0$ & -0.65 \\
 \hline
 $0.25$ & -0.66  \\
 \hline
 $0.5$ & -0.67 \\ 
 \hline
  $0.75$ & -0.69 \\ 
  \hline
   $0.9$ & -0.7 \\ [1ex]
 \hline
\end{tabular}
 \caption{Best-fit late-time luminosity decay indices, in the case of finite ISCO stress, for different black hole spin.}
 \label{table1}
 \end{table}

\subsubsection {Kerr disc, Finite ISCO stress, $W^r_\phi \propto r^{\mu}$}

The general agreement between our numerical solutions and the observational decay indices for the case of finite ISCO stress is not a special feature of a particular $r$-dependence of the turbulent stress.  Table (\ref{table2}) shows the best-fit decay indices for a rapidly rotating black hole, $a/r_g = 0.9$, for a number of different turbulent stress profiles, parameterised as a power law by $\mu$, $\W \propto r^\mu$.  The analytic value of $n$ from equation (\ref{time_dependance}) is appropriate to a Green's function superposition of negative index Bessel functions.   (See Appendix A1 for further details).   The true solution is somewhat more complicated, and thus only rough agreement is expected.  These results nevertheless serve to demonstrate that the exact radial dependence of the turbulent stress is hardly crucial for general agreement between numerical and observational results. This robustness is important, as the particular form of the turbulent stress is not known in detail, apart from an expectation that the stress is likely to fall-off gradually with distance.

 \begin{table}
\centering
 \begin{tabular}{||c | c | c||} 
 \hline
~Stress Index $\mu~$ & $n~$(numerical) & $n~$ (eq. \ref{time_dependance})  \\ [0.5ex] 
 \hline\hline
  $1/2$ & $-0.45$ & $-0.5$ \\ 
  \hline
 $0$ & $-0.61$ & $-0.67$ \\ 
 \hline
 $-1/2$&  $-0.70$ & $-0.75$ \\
 \hline
 $-1$ & $-0.76$ & $-0.8$ \\
 \hline
 $-3/2$ & $-0.78$ & $-0.83$  \\
 \hline
 $-2$ & $-0.80$ & $-0.86$ \\ [1ex]
 \hline

\end{tabular}
 \caption{Comparison between late time numerically determined luminosity decay indices and the analytical form eq.\  (\ref{time_dependance}).  $\W$ is assumed to vary as $r^{\mu}$ (therefore finite at the ISCO), and $a/r_g=0.9$ here, rapid rotation. }
 \label{table2}
 \end{table}

\subsubsection {Kerr disc, Vanishing ISCO stress}
\begin{figure}
	\centering
           \includegraphics[width=7.5cm,clip=true, trim=0cm 0cm 0cm 0cm]{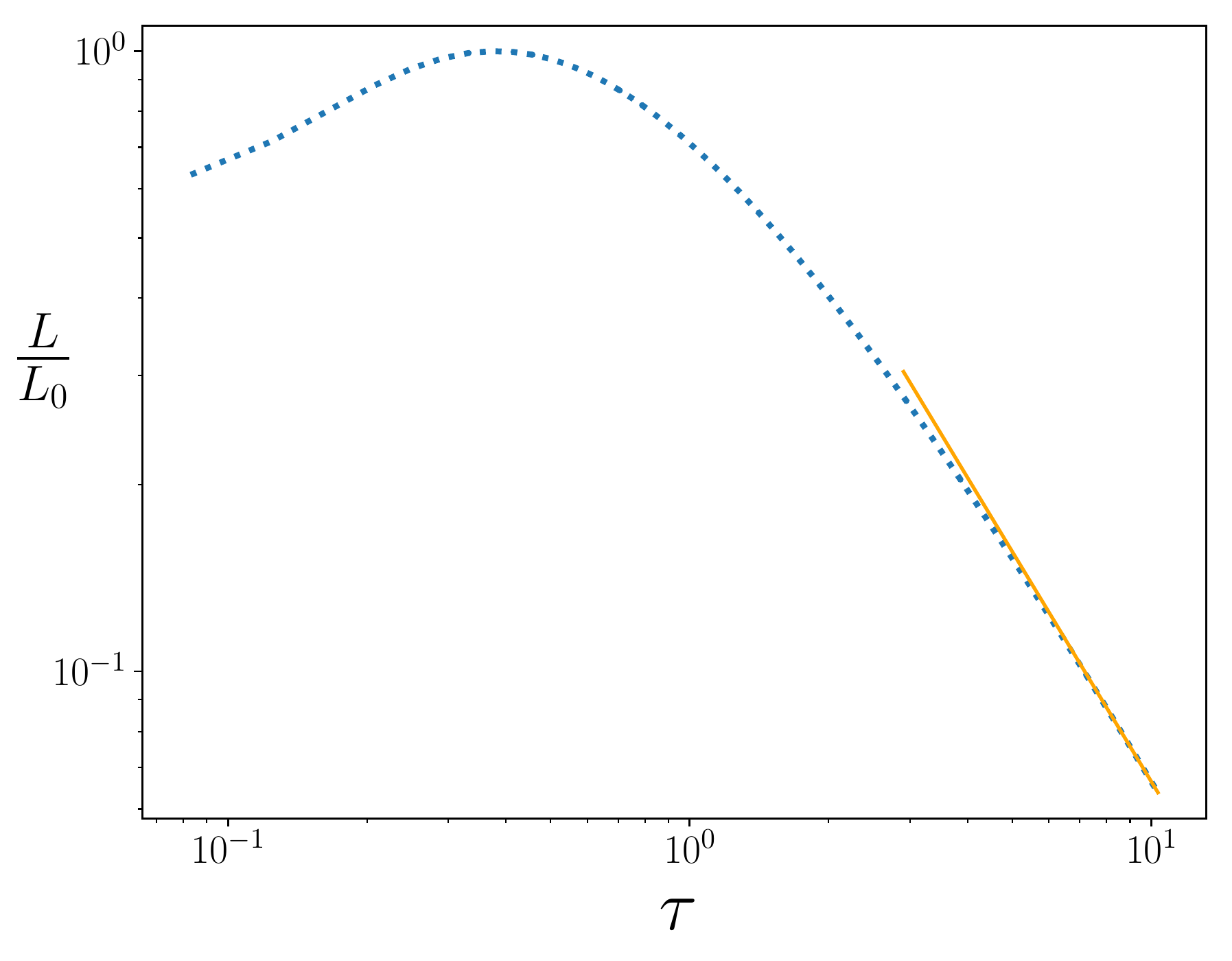}
	\caption{The Luminosity $L(t)$ for the case of vanishing stress at the ISCO, based on solving the full Kerr equation (\ref{27q}).
	The fitted late time behaviour $L\sim t^{-1.23}$, agrees very closely with the Newtonian solution, $L\sim t^{-5/4}$. The dimensionless time $\tau$ as defined in eq.\  (\ref{tau}) is plotted on the x-axis.}
\label{vanishingprofile}
\end{figure}

For this numerical model, the stress vanished near the ISCO ($r_I$) as $W^r_\phi \propto (r - r_I)^2$ near the ISCO (Balbus 2017).   This behaviour continues through $r = 10r_g$, at which point it switches piecewise continuously to $W^r_\phi \propto r^{-1/2}$ for $r > 10r_g$.  The luminosity profile $L(t)$ for a Schwarzschild black hole ($r_I = 6r_g, a = 0$) is displayed Figure (\ref{vanishingprofile}).  

The best-fit decay index for vanishing ISCO stress, $n = -1.23$,  is sharply different from all of the non-vanishing stress cases, but very similar to the Cannizzo et al.\ (1990) canonical Newtonian value of $-19/16\simeq -1.19$. The steeper fall-off is sharply discrepant with the observations showing a distinctly more shallow fall-off with time (Table [\ref{tab3}]).   This suggests that in determining the late-time luminosity behaviour both the stress boundary condition and relativistic dynamics are of importance, neither one alone reproduces the observational results.

\subsubsection{Summary of numerical results}

We have numerically integrated the time-dependent solutions of equation (\ref{27q}), the thin disc equation for the evolution of a turbulent disc.   Two different inner boundary conditions were explored, that of vanishing stress at the ISCO and finite stress.   For the former,  $\zeta$ vanishes at the ISCO, while for the latter, the radial gradient of $\zeta$ vanishes.   This produces two very distinct late time behaviours for the integrated luminoisty $L(t)$, with the finite stress solutions falling of less rapidly than $t^{-1}$ and the vanishing stress solutions more rapidly than $t^{-1}$.    The latter are in good accord with earlier Newtonian disc calculations, but the former are in better accord with observations.     The dichotomy of luminosity falling more steeply than $t^{-1}$ for vanishing ISCO stress and less rapidly than $t^{-1}$ for finite ISCO stress persists with different hole angular momenta, and is insensitive to stress behaviour nonlocal to the ISCO.    

It has been argued that vanishing stress at the ISCO may be expected on the grounds of stability (Balbus 2017), but clearly this question needs to be revisited!  

\subsection {Analytic modal solutions of the reduced equation}

We may understand the key features of our numerical solutions, such as the dichotomy of $L(t)$ power law fall-offs, by examining the Laplace modes of a reduced disk model.   This involves adopting a simple piecewise continuous form for $U'_\phi$.

\subsubsection {Piecewise continuous form for $U'_\phi$}
The Schwarzschild metric angular frequency $\Omega$ is given by
\beq
\Omega={d\phi\over dt} ={U^\phi\over U^0} = \sqrt{r_g\over r^3},
\eeq
and the angular momentum gradient is
\beq
 U'_\phi =  {\Omega\over 2} {r-6r_g\over (1-3r_g/r)^{3/2}} = {\sqrt{r_g}\over 2} {r-6r_g\over (r-3r_g)^{3/2}} .
\eeq
The angular momentum gradient $U'_\phi$ vanishes at the ISCO radius $r_I=6r_g$ and approaches $r\Omega/2$ in the Keplerian zone, $r\gg r_g$.    For analytic purposes, we shall model $U'_\phi$ near the ISCO by its local linear form,
\beq\label{inn}
U'_\phi = \sqrt{2} \Omega_I (r-r_I)\equiv   \sqrt{2} \Omega_I x\quad {\rm {(ISCO)}}
\eeq
where $\Omega_I =\sqrt{r_g/r_I^3}$ and $x=r -r_I$.   In the outer Keplerian disc, we have as usual
\beq\label{out}
U'_\phi = {r\Omega\over 2} = {1\over 2}\sqrt{GM\over r}\quad  {\rm {(Kepler)}}
\eeq
Our model angular momentum gradient, denoted $u'_\phi$, will be the piecewise continuous compilation of equations (\ref{inn}) and (\ref{out}).   Extrapolating (\ref{inn}) to the point where it matches (\ref{out}) at the radius denoted $r_m$, we have
$$
u'_\phi = \sqrt{2} \Omega_I (r-r_I) \quad (r<r_m)
$$
\beq
\quad u'_\phi = {r\Omega\over 2} \quad (r\ge r_m)
\eeq
The matching radius $r_m$ is determined by continuity of $u'_\phi$ at $r=r_m$, the radius at which the two curves cross:
\beq
r_m/r_g = 3(3+\sqrt{5})/2 = 7.854,
\eeq
which is slightly beyond the ISCO radius $6r_g$.  This gives
$$x_m/r_g= {3}(\sqrt{5}-1)/2= 1.854 .$$  
Here, for the sake of simplicity, we shall assume constant $\W$, denoted in the model as $w$.   (With dimensions of length $\times$ velocity$^2$,  a self-similar Keplerian stress tensor $\W$ would in fact lead to a constant-with-$r$ scaling.)    This obviously implies a {\em nonvanishing} ISCO stress;  we consider the case of vanishing ISCO stress in \S 2.6.2.   Our formal reduced equation takes the form 
\beq\label{toy2}
{\dd y \over \dd t} =  {w}{\dd\ \over \dd r}\left[ {1\over u'_\phi}   {\dd y \over \dd r}\right].
\eeq
For $r<r_m$ this becomes:
\beq\label{toy3}
{\dd y \over \dd t} =  {w\over \sqrt{2}\Omega_I}{\dd\ \over \dd x}\left[ {1\over x }   {\dd y \over \dd x}\right].
\eeq
while for $r\ge r_m$ the equation is 
\beq\label{toy4}
{\dd y \over \dd t} =  {2w\over \sqrt{GM}}{\dd\ \over \dd r}\left[ r^{1/2} {\dd y \over \dd r}\right].
\eeq
The joining boundary condition is continuity of $y$ and $dy/dr$ at $r=r_m$.

\begin{figure}
	\centering
           \includegraphics[width=7.5cm,clip=true, trim=0cm 0cm 0cm 0cm]{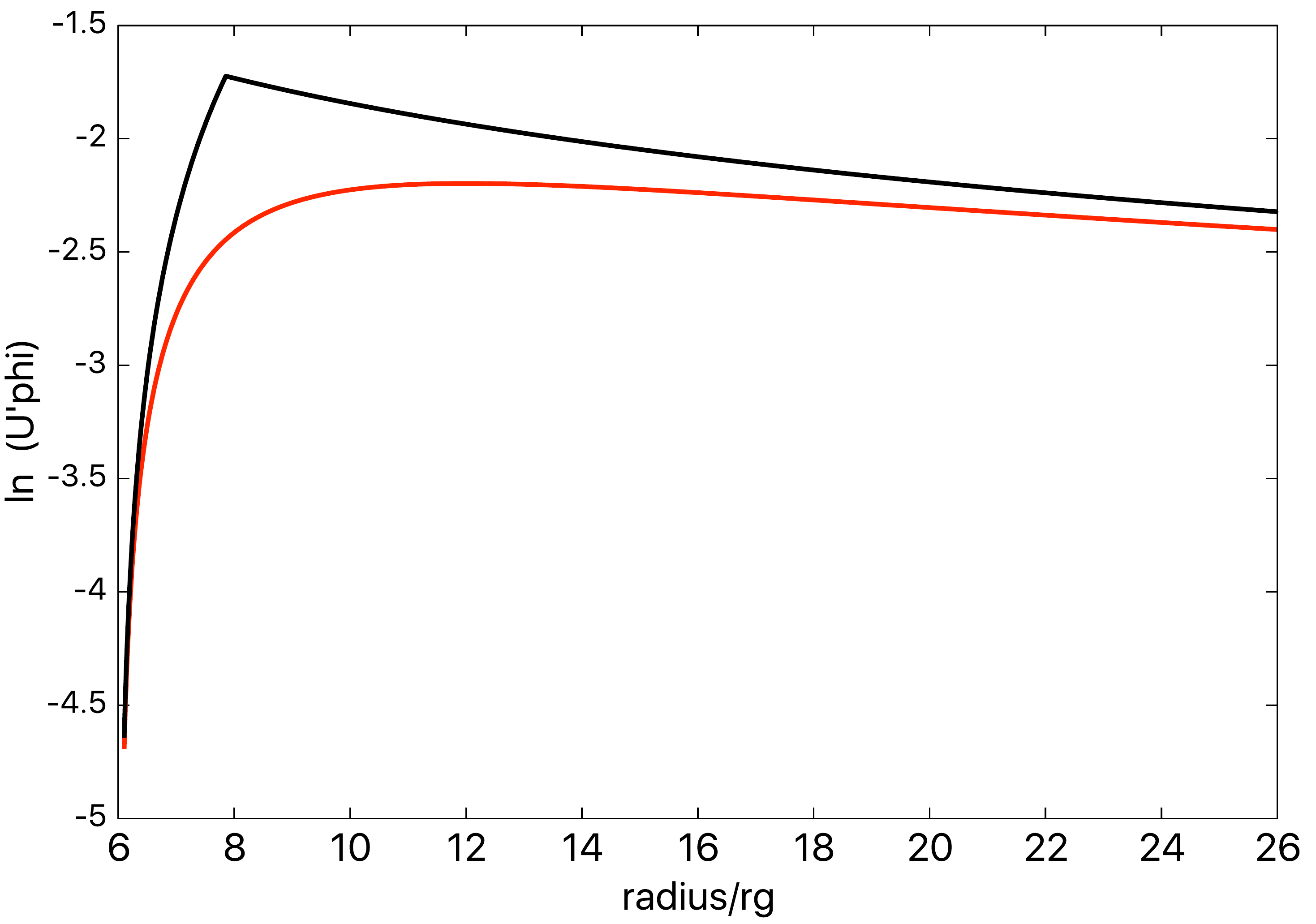} 				
	\caption{$\ln U'_\phi$ versus $r/r_g$.  Piecewise continuous fit (in black) to exact Schwarzschild equation (9) (in red), based on the ISCO approximation (10) for $r\le r_m$, and the Keplerian form (11) for $r\ge r_m$.  $U'_\phi$ calculated with $r_g=1$.} 
\label{fit}
\end{figure}

\subsubsection{Laplace Modes}
We seek stable solutions with time dependence $e^{-st}$, $s\ge 0$, i.e., Laplace transforms.   The solution for a particular initial condition is then a superposition of such (mathematically complete) modes.   The reduced ISCO equation (\ref{toy3}) now reads 
\beq\label{toy5}
-{s\sqrt{2} \Omega_I\over w} y  = {d \over d x}\left[ {1\over x }   {d y \over dx}\right].
\eeq
The desired stable mode that decays exponentially inside the ISCO ($x<0$) and is oscillatory for $x>0$ is:
\beq\label{isco}
y= {\rm Ai}'(-kx), \qquad k^3 = {s\sqrt{2} \Omega_I\over w},
\eeq
as noted in Balbus (2017) for this same problem.   Here Ai is the usual Airy function (Abramowicz \& Stegun 1965) and ${\rm Ai}'$ the derivative with respect to its argument.  
This solution is valid for $r_I\le r \le r_m$, and in effect constitutes the appropriate finite stress inner boundary condition: the exterior, Newtonian domain solution must join smoothly onto the inner soultion (\ref{isco}).  Note that this
condition is different to that of Riffert (2000), who set the density at the ISCO to zero by way of imposing a vanishing stress boundary condition. 

For $r \ge r_m$, the Keplerian form (\ref{toy4}) is appropriate.  Writing this in terms of $\xi=\sqrt{r}$ leads to:
\beq\label{newt}
-{s\sqrt{GM}\over 2w} \xi y = {d^2 y\over d\xi^2} , \qquad \xi\equiv r^{1/2}.
\eeq
This equation has two linearly-independent solutions, $y_+$ and $y_-$, where
\beq\label{c1c}
y_\pm = \sqrt{\xi} J_{\pm 1/3} \left({2\alpha \over 3}\xi^{3/2}\right), \quad \alpha^2 = {s\sqrt{GM}\over 2w},
\eeq
where $J_{1/3}$ and $J_{-1/3}$ are standard Bessel functions of order $1/3$ and $-1/3$ respectively.  
In general, the outer Kelplerian zone solution will take the form of a linear superposition
\beq\label{c1c2}
y = C_1 y_- + C_2 y_+,
\eeq
where $C_1$ and $C_2$ are constants determined by joining this outer Keplerian solution and its first derivative smoothly to the inner solution (\ref{isco}) at the matching radius $r_m$.    In general these two constraints can be satisfied only by a superposition of both $y_+$ and  $y_-$, and {\em it is this admixture that determines the late time behaviour of the general solution. }   

\subsection {Newtonian versus Kerr discs}

In a strictly Newtonian system, equation (\ref{toy4}) has only one solution that leaves $\Sigma$ finite as $r\rightarrow 0$, the solution $y_+$
\beq\label{ynw}
y_+ = \sqrt{\xi} J_{1/3} \left({2\alpha \over 3}\xi^{3/2}\right),
\eeq
which vanishes linearly in $r$ in this limit.    The superposition integral (Gradshteyn \& Ryzhik 2014)
$$
\int^\infty_0 J_p(\sqrt{s}X)J_p(\sqrt{s}X_0)  e^{-st}\, ds = \qquad\qquad\qquad\qquad\ \ \ \ \ \ \ \ 
$$
\beq\label{G2}
\ \ \ \ \ \ \ \ \ \  \qquad {1\over t}\exp\left(-X^2-X_0^2\over 4t\right)I_p\left(XX_0\over 2t\right),
\eeq
tells us how construct the Green's function for $J_p$ modes.  
Here $p$ is any complex number whose real part exceeds $-1$, and $I_p$ is the standard modified Bessel function.   The right side is proportional to a delta function $\delta(X-X_0)$ as $t\rightarrow 0$, a result that is independent of $p$.  Note that while we can always multiply $y_+$ by an arbitrary function of $s$ and retain a true mode, the particular Green's function superposition (\ref{G2}) requires a Bessel function of the form shown in (\ref{ynw}) with no $s$-dependent coefficient.   

The late time $t\rightarrow \infty$ asymptotic behaviour of the $I_p$ function leads to a time dependence in the surface density scaling as 
\beq
\Sigma \sim {1\over t^{1+p}}
\eeq 
As this is the only quantity that varies with time in the emission integral, this is also the late time power law time dependence of the total luminosity $L$.   The leads to a $t^{-4/3}$ dependence in our simple constant angular momentum stress model ($p=1/3$), but it is not very sensitive to how the stress is modelled.   A constant viscosity model leads to $t^{-5/4}$ (Pringle 1981), while a more complicated $\alpha$-model produces values near $-19/16\simeq -1.19$ (Cannizzo et al.\ 1990).      Neither of these is can be comfortably stretched to the observed peak near $-0.75$ in the late time histogram of values given by Auchettl et al. (2017).    

An initial $t=0$ delta function superposition of $J_{-1/3}$ Bessel functions, on the other hand, would produce a late time luminosity scaling $t^{-2/3}$, much closer to the observed AGR histogram peak.
(A constant viscosity model at $t^{-3/4}$ is yet better.)    The apparent difficulty is that these modes lead to singular behaviour in the surface density (and perhaps the mass accretion rate) as $r\rightarrow 0$.    

In reality, the inner disc of a Kerr black hole is cut-off at the ISCO.  The $r=0$ singularity is not relevant to this type of calculation.    Even in Newtonian modelling some inner disk cut-off is required to avoid a singularity in the accreted matter.   But if $r=0$ is actually outside the domain, we should expect a superposition of $J_{1/3}$ and $J_{-1/3}$ solutions to be present in the bulk of the disc.   This in itself is not enough to guarantee late time $t^{-2/3}$ luminosity behaviour; it depends upon the precise $s$-dependence of our modal superposition.   But we now understand how in principle the more shallow late time fall-off that emerges from direct numerical integration of equation (\ref{27q}) arises.

\subsection{Construction of solution}
\subsubsection {Finite ISCO stress}
We return now to the task of finding the matching coefficients $C_1$ and $C_2$ of the outer Keplerian zone solution by requiring continuity of $y$ and its first derivative $dy/dr$ at the matching radius $r_m$.   For future reference we note, 
\beq
{d {\rm Ai}'(-kx)\over dr} ={d {\rm Ai}'(-kx)\over dx} = -k {\rm Ai}''(-kx) = k^2 x {\rm Ai}(-kx)
\eeq
where the prime $'$ denotes differentiation with respect to the displayed functional argument and the differential equation for the Airy function has been used in the final equality.     Also, with $y$ given by equations (\ref{c1c}) and (\ref{c1c2}), standard Bessel function identities allow us to write
\beq
{dy\over dr} = {\alpha\over 2}  \left[ - C_1 J_{2/3}\left({2\alpha \over 3}r^{3/4}\right)+C_2 J_{-2/3}\left({2\alpha \over 3}r^{3/4}\right)\right]
\eeq
To avoid a cluttered appearance in the equations, let us define
\beq
\tilde\alpha = 2r_m^{3/4} \alpha/3.
\eeq
The matching conditions for $y$ and $dy/dr$ at $r=r_m$ are then
\beq\label{s1}
{{\rm Ai}'(-kx_m) \over r_m^{3/4}} = C_1J_{-1/3}(\tilde\alpha) + C_2J_{1/3}(\tilde\alpha)
\eeq
\beq\label{s2}
{2k^2 x_m\over \alpha} {\rm Ai}(-kx_m) = -C_1 J_{2/3}(\tilde\alpha) + C_2 J_{-2/3}(\tilde\alpha)
\eeq
Solving for $C_1$ and $C_2$:
\beq
C_1={ {\rm Ai}'(-kx_m) \ J_{-2/3}(\tilde\alpha)\over r_m^{3/4}\ {\rm Wr}} - {2x_mk^2 {\rm Ai}(-kx_m)\  J_{1/3}(\tilde\alpha)\over \alpha\ {\rm Wr}}
\eeq
\beq
C_2={ {\rm Ai}'(-kx_m) \ J_{2/3}(\tilde\alpha)\over \ r_m^{3/4}{\rm Wr}}  +{ 2x_m k^2{\rm Ai}(-kx_m) \  J_{-1/3}(\tilde\alpha)\over\alpha{\rm Wr}}
\eeq
where ${\rm Wr}$ is the Wronskian (Abramowitz \& Stegun 1965):
\beq
{\rm Wr}\equiv J_{1/3}(\tilde\alpha)J_{2/3}(\tilde\alpha) + J_{-1/3}(\tilde\alpha)J_{-2/3}(\tilde\alpha) = {3\sqrt{3}\over 2\pi\alpha r_m^{3/4}}.
\eeq
The late time ($t\rightarrow \infty$) superposition of $e^{-st}$ modes will be dominated by contributions from small $s$.   In this limit, it is the ${\rm Ai}' J_{-2/3}$ term in the $C_1$ numerator that is dominant ($\sim s^{-1/3}$), and therefore so is $C_1$ itself.   When the inner ISCO boundary hosts a nonvanishing stress, the late time behaviour of the matching outer Keplerian zone is therefore dominated by the $J_{-1/3}$ modes.   This is precisely what we would expect given a late time power law of $-2/3$ for the luminosity and the (\ref{G2}) superposition integral with $p=-1/3$.  

\subsubsection {Vanishing ISCO stress}
If the stress tensor vanishes at the ISCO, then the local solution near $x=0$ is (Balbus 2017):
\beq
y = xJ_2(2\sqrt{\beta x}), \quad \beta = 2s\sqrt{2}\Omega_I/\W
\eeq
where $J_2$ is the Bessel function of order $2$.  Using (Abramowitz \& Stegun 1965):
\beq
{dy\over dr} = \sqrt{\beta x} J_1(2\sqrt{\beta x})
\eeq
The system of equations to be solved now is (suppressing the Bessel function arguments on the right side of the equations):
\beq
{x_m J_2 (2\sqrt{\beta x_m})\over r_m^{3/4}} = C_1J_{-1/3} + C_2J_{1/3}
\eeq
\beq
{2\sqrt{\beta x_m} J_1(2\sqrt{\beta x_m})\over \alpha}  = -C_1 J_{2/3} + C_2 J_{-2/3}
\eeq
This is identical to the system (\ref{s1}) and (\ref{s2}) with $x_m J_2$ replacing ${\rm Ai}'$
and $\sqrt{\beta x_m} J_1$ replacing $k^2x_m {\rm Ai}$.   When not shown, the argument of all integer Bessel functions is understood to be $2\sqrt{\beta x_m}$; as before, fractional Bessel functions all have argument $2r_m^{3/4}\alpha /3$.  The solution of this system is
\beq
C_1={x_m J_2\ J_{-2/3}\over {\rm Wr}\  r_m^{3/4}} - {2\sqrt{\beta x_m}J_1\   J_{1/3}\over {\rm Wr} \ \alpha}
\eeq
\beq
C_2={1\over {\rm Wr}}\left[ {x_m J_2 \ J_{2/3}\over r_m^{3/4}} +{2\sqrt{\beta x_m}J_1\  J_{-1/3}\over \alpha}\right]
\eeq
The small $s$ scalings are
\beq
\alpha\sim s^{1/2},\   \beta\sim s, \  J_p\sim s^{p/2}\quad {\rm (for\ all}\ p{\rm )}.
\eeq
Now the dominant term for small $s$ is the $J_1 J_{-1/3}$ group in the $C_2$ coefficient.   This in turn means that the dominant contributing modes in the Keplerian zone are {\it positive} indexed, $J_{1/3}$ modes.  Once again, this is just what is expected on the basis of a (\ref{G2}) modal superposition integral with $p=1/3$ and from the observed late time $t^{-4/3}$ time dependence found in the numerical simulations.

\begin{table}
\centering
\begin{tabular}{| l | c |}
\hline
ASASSN-14li & $-1.0$ \\ \hline
 Swift J1644+57 & $-0.71$ \\ \hline
Swift J2058+05& $-0.16$ \\ \hline
 XMMSL1 J0740-85 & $-0.75$ \\ \hline 

\end{tabular}
\caption{ The four well-observed sources from AGR (left) and their deduced late time luminosity power law index (right).}
\label{tab3}
\end{table}

\section {Discussion}

Figure (15) of AGR shows a histogram of power law indices of X-ray selected TDEs.   The curves have been separated by early (solid lines) and late time (dashed lines) divisions, as well as whether the event is a strong TDE candidate (shown in blue) or only ``likely.''  The dominant peak in the histogram is for late time, likely X-ray TDEs, and it occurs for a power law index of $n\simeq- 0.75$.  In fact, the histogram is somewhat schematic because there are only four well-observed late time confirmed TDEs in the AGR sample.   These, together with their inferred power law index, are listed in Table 3.   While errors in these values are somewhat difficult to assess, one significant figure is probably a reasonable working assumption, and what is therefore striking from this table is that none of these indices is larger than 1 (in magnitude).    This accords nicely with our own numerical findings that Keplerian discs joining onto an inner, finite stress, ISCO region  also do not have late time power law fall-offs in luminosity steeper than one.    While it is premature to conclude that all late time TDEs have settled into an accreting thin disc (there are too many ways for a ``train wreck'' to unfold), there seems to be a case that at least some of them may well be.   It is striking and gratifying that the classical Newtonian results may be recovered and that new solutions can be achieved from our approach, and that the latter offer a new theoretical route to understanding the shallow power law luminosity fall-offs with completely conventional disc physics.    We also remark in passing that, compared with direct interpretation of disc spectra, the late time temporal behaviour of an evolving disc is a more powerful, less ambiguous, discriminator for the presence or absence of finite stress at the ISCO.  The disadvantage is of course that the best time for observing this is when the source is faintest.

As we have noted, the question of whether the turbulent stress must vanish at the ISCO has been controversial; the view that it must vanish on dynamical grounds has retained prominent advocates (e.g.\ Paczy\'nski 2000).   Moreover, one of the current authors argued in an earlier paper (Balbus 2017) that a vanishing ISCO stress was to be expected on the grounds of greater stability.   But simulations often show magnetic stress remaining finite down to the ISCO (e.g. Noble, Krolik, \& Hawley 2010), and there is physical basis for understanding why an outward angular momentum flux constant should be present when the disc flow sharply transitions from rotational dominance to inward streaming (Agol \& Krolik 2000).    That a finite ISCO stress may in fact exhibit some degree of flow instability from the tunnelling of unstable modes from within the ISCO radius need not be a basis for rejection:  this sort of behaviour in the region outside the ISCO is liable to be little more than orbital inspiral before turning to true plunging, once the ISCO is crossed.   Indeed, something very much like this behaviour is observed in detailed numerical MHD simulations.   The fundamental content of our analysis is likely to be preserved even with inspiraling near the ISCO, a claim that may now checked by combining our semi-analytic approach with controlled 3D MHD simulations.   In short, there seems to be nothing particularly unphysical about finite magnetic ISCO stresses.  TDE light curves may well be a powerful observational constraint, if they consistently show late time power law indices less than unity.   

Finally, we reiterate that discs are much more complicated than our 1970's era thin disc model.   Real discs need not be thin;  they have outflows, jets, coronae, and as yet poorly understood major state transitions.   The main point, however, is that four decades after its inception, even the ordinary thin disc model has not been understood in all of its temporal manifestations.   This is not just bookkeeping; at least some observations seem to be quite well fit by simple thermal modelling!     We cannot hope to understand with any depth, or assess the need for, more complex calculations without a better understanding of our baseline modelling.    Perhaps the simple solutions discussed here are revealing behaviour which will allow us to understand some of the interesting temporal features of a class of TDEs.    The mathematical tools are now in place for studying evolving relativistic discs.

\section*{Acknowledgements}
It is a pleasure to acknowledge useful conversations with K.\ Auchettl, R.\ Fender,  J.\ Guillochon, K.\ Horne, P.\ Ivanov, W. Kley, J.\ Krolik, C.\ McKee, and E.\ Ramirez-Ruiz.     Comments from our referee have improved the presentation.    SAB acknowledges support from the Royal Society in the form of a Wolfson Research Merit Award, and from STFC (grant number ST/N000919/1).

\section*{A1: Solution for Keplerian power law stress tensor}
We present here the solution of equation (\ref{toy4}) assuming that $\W$ behaves as a power law in $r$ throughout the Keplerian zone:
\beq
\W=w_m(r/r_m)^\mu
\eeq
where $\mu$ and $w_m$ are constants.   (We shall assume that $\W=w_m$ within the matching zone, remaining constant down to the ISCO.)  

\beq\label{toy5}
{\dd y \over \dd t} =  {2w_mr^\mu\over r_m^\mu\sqrt{GM}}{\dd\ \over \dd r}\left( r^{1/2} {\dd y \over \dd r}\right).
\eeq
With $\xi=r^{1/2}$ as in \S 2, this may be written
\beq
{\dd y \over \dd t} =  {w_m\over 2r_m^\mu \sqrt{GM}}\  \xi^{2\mu -1} \   {\dd^2 y \over \dd \xi^2}.
\eeq
We seek the Laplace modes with time dependence $e^{-st}$.  With $y$ regarded as the Laplace amplitude,
the equation becomes
\beq
{d^2 y \over d \xi^2} =   - {2sr_m^\mu \sqrt{GM}   \over w_m}\  \xi^{1-2\mu }\ y\equiv - s\gamma^2 \xi^{1-2\mu } y,
\eeq
where 
\beq
\gamma^2=2r^\mu_m\sqrt{{GM\over w^2_m}}.
\eeq
The solution to this equation is:
\beq
y= r^{1/4} J_{\pm {1\over 4q}}\left( s^{1/2}\gamma r^q \over 2q\right), \quad q={3-2\mu\over 4}
\eeq
With $\xi =r^{1/2}$, we recover (\ref{c1c}) in the limit $\mu\rightarrow 0$.   If the inner boundary condition at $r=0$ requires the vanishing of $y$, the positive index solution should be selected; otherwise both solutions are valid and should be retained throughout their region of validity.    

The superposition of either one of these distinct solutions via a Laplace integral of the form (\ref{G2}) leads to the respective Green's function solutions:
\beq
y = { r^{1/ 4}\over t} \exp \left[ -{\gamma^2(r^{2q}+r_0^{2q})\over 16 q^2 t} \right] I_{\pm{1\over 4q}}\left({ \gamma^2 r^q r_0^q\over  8 q^2 t}\right)
\eeq
for a ring initially laid down at $r=r_0$.   
This, in turn, leads to late time luminosity behaviours of the form
\beq
L(\tau) \propto  \tau^{-(1\pm 1/4q)},
\eeq
where we have normalised the time via the dimensionless variable $\tau$ (Pringle 1981):
\beq
\tau = {16 q^2 t\over \gamma^2}.
\label{tau}
\eeq
Consider first $I_{1\over 4q}$, which is appropriate to a vanishing ISCO stress solution.   This leads to 
\beq
L(\tau) \propto \tau^{-(4-2\mu)/(3-2\mu)}.
\eeq
For declining outward stress, $\mu\le 0$,  this is always larger than unity - as found in our vanishing stress Newtonian calculations.   Taking the $-1/4q$ solution leads, on the other hand, to 
\beq
L(\tau) \propto  \tau^{-(2-2\mu)/(3-2\mu)},
\label{time_dependance}
\eeq
a power law index always {\em less} than unity.  This result is in better accord with observations.  If we are indeed viewing the late stages of accreting discs in TDEs, it suggests a significant late time admixture of solutions dominated by $J_{-1/4q}$ modes and a finite ISCO stress.   This, in turn, is in good agreement with the $s\rightarrow 0$ analysis of section 2.6.2.

\section*{A2: Numerical Method}
With $Q$ defined by equation (\ref{dQ}), the fundamental equation (\ref{fund}) may be written:
\beq
{\dd\zeta\over \dd t} =  {e^Q \W\over U^0}{\dd\ \over \dd r}{e^{-Q}\over U'_\phi} \left[   {\dd\zeta \over \dd r}\right].
\eeq
Recall that we work in `Boyer-Lindquist' co-ordinates in their near-equator form: $t$ is time, as measured at infinity; $r$ is cylindrical radius; $\phi$ is azimuthal angle, and $z$ is height above equator.  The line element is given by 
\begin{multline}
ds^2 = -\left(1 - \frac{2r_g}{r}\right)  \text{d}t ^2 - \frac{4r_g a }{r}~ \text{d} t ~\text{d} \phi \\ + \frac{\text{d} r ^2}{1 - {2r_g}/{r} + {a^2}/{r^2}}  
  + \left( r^2 + a^2 + \frac{2r_g a^2 }{r} \right) \text{d} \phi ^2 + \text{d}z^2
\end{multline}
The circular orbit solutions in the equatorial plane are given by (e.g., Hobson et al.\ 2006):
 \begin{align}
U_0 &= -\frac{1-2r_g/r +a\sqrt{r_g/r^3}}{\left( 1- 3r_g/r + 2a\sqrt{r_g/r^3}\right)^{1/2} }\\
  U^0 &= \frac{1+a\sqrt{{r_g}/{r^3}}}{\left({1 - {3r_g}/{r} + 2a\sqrt{{r_g}/{r^3} } }\right)^{1/2}} \\
 U_\phi & = \sqrt{r_g r}\ \frac{1 + {a^2}/{r^2} - 2a\sqrt{{r_g}/{r^3}}}{\left({1 - {3r_g}/{r} + 2a\sqrt{{r_g}/{r^3} } }\right)^{1/2}} \\
 U^\phi &= \frac{\sqrt{{r_g}/{r^3}}}    {\left( 1 - {3r_g}/{r} + 2a\sqrt{{r_g}/{r^3} } \right)^{1/2}} \\
 \Omega &= \frac{U^\phi}{U^t} = \frac{\sqrt{{r_g}/{r^3}}}{1 + a\sqrt{{r_g}/{r^3}} }
 \end{align}
Using $dQ/dr\equiv-U_\phi U^\phi d(\ln\Omega)/dr$, direct calculation gives
\beq\label{58Q}
 e^{-Q} = \frac{1+a\sqrt{{r_g}/ {r^3}}} {\left(1 - {3r_g}/{r} + 2a\sqrt{{r_g}/{r^3} }\right)^{1/2} }=U^0,
 \eeq
 and 
 \beq
 U_\phi ' = \frac{\sqrt{r_g} \left( a\sqrt{r_g} + r^{{3}/{2}} \right) \left( r^2 - 6r_g r - 3a^2 + 8a\sqrt{r_g r}\right)}{2r^4 \left(  1 - {3r_g}/{r} + 2a\sqrt{{r_g}/{r^3}}   \right)^{{3}/{2}}}.
 \eeq
Substituting into the full evolution equation, we obtain
\beq\label{60z}
 \frac{\partial \zeta}{\partial t} = \frac{2W^r_{\phi}}{\sqrt{r_g}(U^0)^2}\frac{\partial}{\partial r}\left[ r^{3/2} 
 F(r) \frac{\partial \zeta}{\partial r}\right],
\eeq
where 
\beq
F(r) = \frac{ 1 - {3r_g}/{r} + 2a\sqrt{{r_g}/{r^3}} }  { r - 6r_g  - {3a^2}/{r} + 8a\sqrt{{r_g}/{r}} }.
\eeq

The position of the (apparently singular) ISCO is given by the solution of the equation $r_I^2 - 6r_g r_I - 3a^2 + 8a\sqrt{r_g r_I} = 0$. Numerical integration of the PDE is unstable in the vicinity of this point.   This problem can be addressed by the substitution:
 \beq
\rho = \left( r - 6r_g  - \frac{3a^2}{r} + 8a\sqrt{\frac{r_g}{r}} \right)^2.
\eeq
There is no (simple) analytic expression for the explicit inverse $r=r(\rho)$, but numerically there is no difficulty with this inversion.    Using
\beq
\frac{\partial \zeta}{\partial r} = 2\left(1+\frac{3a^2}{r^2}-4a\sqrt{\frac{r_g}{r^3}}\right)
 \left( r - 6r_g  - \frac{3a^2}{r} + 8a\sqrt{\frac{r_g}{r}} \right)\frac{\partial \zeta}{\partial \rho}
\eeq
we may remove the numerical singularity at the ISCO.   Upon full substitution of $\rho$ for $r$, we have 
 \beq\label{evolutioneqn}
 \frac{\partial \zeta}{\partial t} = W^r_\phi A(\rho) \left[ B(\rho) \frac{\partial \zeta}{\partial \rho} + C(\rho) \frac{\partial^2 \zeta}{\partial \rho^2} \right]
 \eeq
 with 
 \beq
 A =  \frac{1 - {3r_g}/{r} + 2a\sqrt{{r_g}/{r^3}} }{\sqrt{r_g}\left(1 + a\sqrt{{r_g}/{r^3}}\right)^2}
 \eeq
 \beq 
 B =  \frac{6}{\sqrt{r}}\left(r-r_g-8a\sqrt{\frac{r_g^3}{r^3}} - \frac{a^2}{r^2}\left(r - 17r_g\right) - 8a^3 \sqrt{\frac{r_g}{r^5}}\right)
 \eeq
 \begin{multline}
 C = 8 r^{\frac{3}{2}} \left(  1 - \frac{3r_g}{r} + 2a\sqrt{\frac{r_g}{r^3}}\right)
\left( r - 6r_g  - \frac{3a^2}{r} + 8a\sqrt{\frac{r_g}{r}} \right) \\
\left(1+\frac{3a^2}{r^2}-4a\sqrt{\frac{r_g}{r^3}}\right)^2 
 \end{multline}
where $r$ is an implicit function of $\rho$. The numerical solution to this equation was found using the implicit finite difference method, with centred finite difference approximations used for spatial $\rho$ derivatives and a forward difference approximation used for the time derivative (Press et al.\  2007).
 
The derivation of the evolution equation is premised upon small perturbations from circular orbits, and so within the ISCO it will quickly break down. Physically, we expect the fluid elements to quickly spiral into the Kerr hole after crossing the ISCO on a timescale similar to the free-fall time. This was demonstrated by Shafee \textit{et al.\ } (2008), who found laminar flow in full GRMHD simulations  within the ISCO. During this phase the fluid elements release almost no radiation and so barely contribute to the disc's spectra (Penna \textit{et al.\ } 2010).  Numerical integration of equation (\ref{evolutioneqn}) is performed for the region of spacetime outside of the ISCO only ($\sqrt{\rho} > 0$), which is both mathematically self-consistent and physically sensible.  
 
Once the full evolution equation has been solved, the time dependent luminosity is straightforward to calculate.  The local flux from a disc annulus is given by (Balbus 2017) 
 \beq
\mathcal{F} = - \Sigma U^0 W^r_\phi \Omega '
\label{flux}
\eeq
In full detail the luminoisty is rather complicated, but fortunately we are interested here only in the gross, late time behaviour, not the precise spectral distribution.   For this, a simple face-on disc model is more than sufficient, indeed the luminosity is largely from the Newtonian disc region.  We retain the gravitational and kinematic redshift effects, which introduce the ratio of observed to emitted flux,  $\left(U^0\right)^{-2}$, and neglect the photon orbit (``ray tracing'') complications.  The total observed luminosity is then given by 
\beq
L(t) \propto \int\limits_{0}^\infty  \sqrt{g_{rr}g_{\phi\phi}}  {\mathcal{F}\over (U^0)^2}~ \text{d}r
\eeq
This may be written explicitly in terms of $r$:  
\beq
L(t) \propto \int\limits_{r_I}^\infty  \frac{\zeta(r,t) \sqrt{r^2 + a^2(1 + 2r_g/ r)}} 
{  r^{7/2}   \left(1 + a\sqrt{r_g/r^3}\right)^2  \sqrt{1 - 2r_g/r + a^2/r^2}}~\text{d}r
\label{luminosity}
\eeq
We do not include the contribution of the disc with $r < r_I$ to the emitted luminosity, effectively terminating the disc emission at the ISCO.  The integral  (\ref{luminosity}) was performed using a standard Simpson-type algorithm.


\label{lastpage}

\end{document}